\documentstyle[12pt,moriond]{article}
\input{psfig}

\begin{document}
\heading{CFRS follow up with ISO, VLA and HST: the cosmic star formation 
rate as derived from the FIR luminosity density}
\author{F. Hammer $^{1}$, H. Flores $^{1}$ }{ $^{1}$ DAEC, Observatoire de Paris Meudon, 92195 Meudon, France} 

\begin{moriondabstract}
Properties of CFRS field galaxies up to z=1 are discussed. Estimations of
the cosmic star formation rate (SFR) lead to serious
problems if they not account for AGN emissions and for light reemitted at
 IR wavelengths. Deep ISOCAM and VLA photometries on one CFRS field 
have been performed. Multi-wavelength analyses from UV to Mid-IR and hence
to radio allow us to classify sources from their spectral 
energy distributions. This provides an estimation of the FIR luminosity
 density related to star formation. The deduced SFR density is free
 of extinction effects and not contaminated by AGN emissions. About 55$\pm$20\%
of the star formation at z$\le$1 is related to FIR emission. If a non
 truncated Salpeter IMF is adopted, the derived
stellar mass formed from z=0 to z=1 seems too high when compared to
the present day stellar mass.\\
An important fraction (30\%) of the star formation at z=0.5-1 seems to be
related to the rapidly evolving population of compact/Irr galaxies. Larger
systems found at z=1, show a slower evolution of their star
formation properties. 
\end{moriondabstract}

\section{Introduction}
The Canada France Redshift Survey (CFRS) has gathered a complete 
sample of $\sim$
600 field galaxies with $I_{AB}\le$ 22.5 with spectroscopic redshifts. This
sample has been widely used to study galaxies at lookback times up to 10 Gyr. 
Evolution of galaxies has been evidenced: at z=0.85 the comoving densities of
2800\AA~ and [OII]3727 were respectively 4.5$\pm$1  and 8$\pm$4 times 
larger than today (Lilly et al, 1996; Hammer et al, 1997).\\

It is beyond doubt that galaxy evolution is mainly associated to a
decrease of the star formation in field galaxies, since epochs of 
$\sim$ 10 Gyr ago. There were larger amounts of blue galaxies in the past
(Lilly et al, 1995) and the fraction of galaxies showing significant emission 
lines ($W_{0}([OII])\ge$ 15\AA~) increases from 13\% locally to more than 50\% at z$\ge$0.5 (Hammer et al, 1997).
 It has been investigated if AGN can substantially contribute to the 
reported evolution.
The fraction of AGN (Seyfert2) is found to be 7.5$\pm$3.5\% at z$\sim$ 0.5, 
 higher than what is found today ($\sim$ 2\%, see Huchra and Burg, 1992). 
But it is found nearly constant with the redshift, when calculated relatively
 to the population of emission line 
galaxies, and so AGNs cannot
be seen as major contributors to the reported luminosity evolution.\\

 40\% of the field galaxy spectra show evidences for a significant 
 population of A stars. This has been
 derived from continuum indices which are well correlated with $W_{0}(H\delta)$.
It implies that star formation is primarily taking place over long periods
of time (typically $\ge$ 1 Gyr), rather than in short duration, high amplitude 
bursts (Hammer et al, 1997).\\

Having those facts in mind, there are important pending questions:
\begin{itemize}
\item  how fair is the estimation of the SFR derived from UV 
( 2800\AA~or [OII]3727) luminosities?
\item can the extinction affect our view of the galaxy evolution ?
\item  are z$\sim$ 1 galaxies showing the same Hubble type 
distribution than those today ?
\item  does evolution affect all galaxy morphological types in the same way ?
\end{itemize}

In this paper we summarize recent developments made beyond the bulk of the CFRS
study. Recall that the CFRS includes 5 fields of 10'X10' which probe a surface
almost 100 times larger than the HDF, and so it is much more suited for
galaxy studies limited in volume to z=1. The CFRS sample is limited by
$I_{AB}\le$22.5, which is exactly coincident to the rest-frame 
$M_{B}(AB)\le$-20.5 at z=0.93. {\it It implies that the CFRS is basically
a (z$\le$1) volume limited sample for $M_{B}(AB)\le$-20.5 galaxies}. The paper is organised as follows:
\begin{itemize}
\item Section 2: calibration of the SFR based on UV luminosities and its deficiencies.
\item Section 3: mid-IR and radio luminosities from deep ISOCAM and 
VLA observations, derivation of FIR luminosities by interpolations and a 
first estimation of the star formation history which is not affected by
 the extinction.
\item Section 4: morphologies from a deep HST follow up, and 
the major contributors to the star formation evolution.
\end{itemize}

\section{SFR calibration to z=1 and beyond}
Calibration of star formation rate in the optical is limited to z=0.5 when
based on the $H\alpha$ line. Other SFR indicators are the 2800\AA~ 
 the [OII]3727 luminosities which could be used up to z=2 and to z=1.5,
respectively. They are more dependent on the metallicity 
and on the extinction than the $H\alpha$ emission.\\
 
ISMs in distant field galaxies present different properties than that
of local ones (Hammer et al, 1997). Emission-line ratio of HII regions 
in z$\sim$0.5 galaxies showed
higher ionization parameters than those of local HII galaxies. About 
30\% of the z$\ge$0.7 galaxies show clues for
 abundances significantly lower than solar values. 
 The [OII]3727/$H\alpha$ luminosity ratio is different on average in CFRS galaxies than in Kennicutt local ones. Indeed, when applied to distant 
galaxies, the Kennicutt (1992) calibration of the
 SFR from the [OII] luminosity leads to uncomfortably high values of 
stellar mass.\\

More recently
we have got spectra up to 1$\mu$m for a sub-sample of z$\le$0.5 CFRS galaxies.
Figure 1 presents relationships between the different SFR indicators:
while [OII]3727 and  2800\AA~ luminosities correlate fairly well, these SFR
indicators present large dispersions when correlated to $H\alpha$ luminosities.
This suggests that extinction affects considerably UV luminosities, and
hence the determination of the SFR. \\
 
\begin{figure}
  \begin{center}
    \leavevmode
    \psfig{file=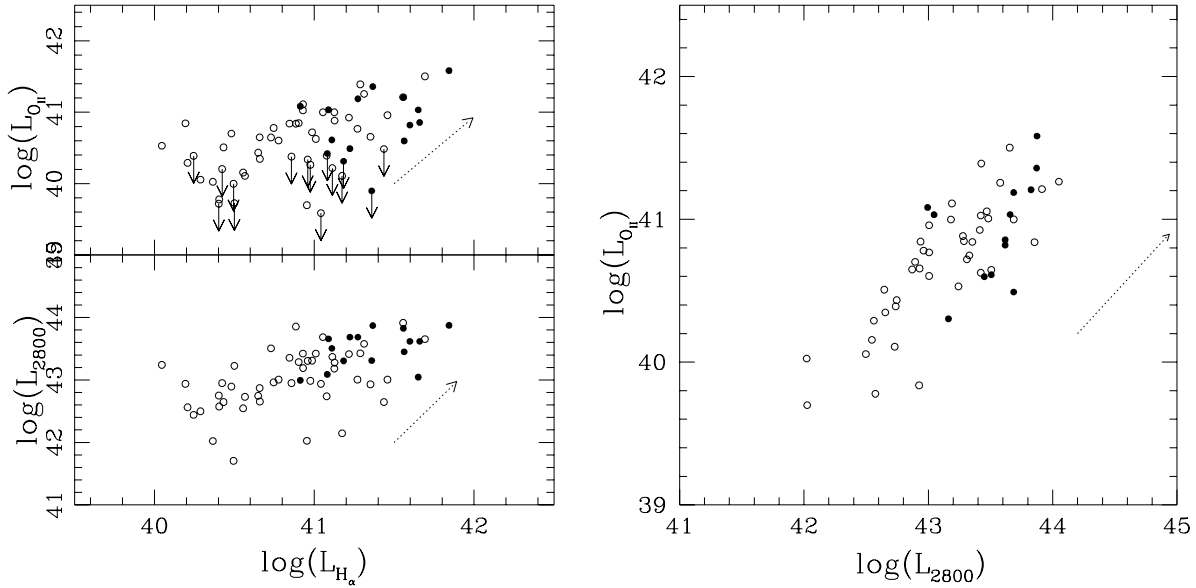,height=8cm,width=16cm} 
  \end{center}
  \caption{{\small {\it (left)} Observed 2800\AA~ and [OII]3727 luminosities 
compared to observed $H\alpha$ luminosities for CFRS galaxies with z$\le$0.5.
    Solid points distinguish $M_{B}(AB)\le$-20.5 and vertical arrows 
indicate an upper limit.
    The dotted line arrow indicate the effect of the extinction correction
 for a standard galactic extinction law. {\it (right)} The relationship between the two observed UV luminosities. Correlation parameter is r=0.7. 
Extinction would have the effect to enhance the dispersion between observed
 UV and $H\alpha$ luminosities. In the right diagram it would move points along
the expected correlation.}}
  \label{figure1}
\end{figure}
\normalsize

\section{FIR luminosity from multiwavelength analysis and SFR density from z=0 to z=1}

This section summarizes two papers by Flores et al (1998a and b).
One CFRS field (1415+52) has been deeply imaged with the Infrared 
Space Observatory (ISO) using ISOCAM at 6.75$\mu$m and 15$\mu$m, with integration times from 1100 to 1200 sec $pixel^{-1}$. 
Careful data analysis and comparison to deep optical (B, V and I),
near-IR (K) and radio (1.4 and 5 GHz) data have allowed us to generate
a catalog of 78 15$\mu$m sources with identifications. 22 redshifts of
galaxies with $I_{AB}\le$ 22.5 are
available in the CFRS database. They have higher
median redshift (Figure 2) and are redder than other field galaxies. Almost all
the star forming galaxies present evidences for an A star population.\\

\begin{figure}
  \begin{center}
    \leavevmode
    \psfig{file=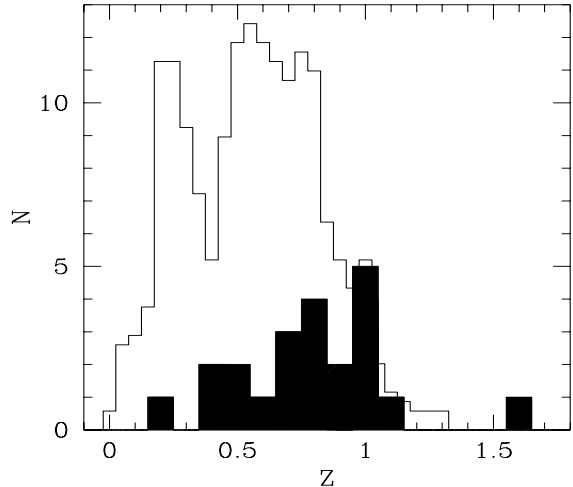,height=7cm,width=8cm} 
  \end{center}
  \caption{ {\small The shaded histogram shows the redshift 
  distribution of the 22 15$\mu$m sources 
   with spectra found in the CFRS database. The non-shaded histogram 
shows the CFRS redshift distribution after 
  rescaling.}}
 \label{figure2}
\end{figure}
\normalsize
   
Source densities are comparable at 6.75$\mu$m (1944 S$>$150$\mu$Jy sources
per square degree, Flores et al, 1998a), 15$\mu$m (2808 S$>$250$\mu$Jy 
sources per square degree, Flores et al, 1998b) and  5 GHz (1440 
S$>$16$\mu$Jy sources per square degree,
Fomalont et al, 1991). Star-forming objects contribute respectively,
50\%, 73\% and 26\% of the extragalactic counts at 6.75$\mu$m,
15$\mu$m and 5 GHz. This suggests that the 60$\mu$m luminosity
density is strongly dominated by emissions related to star formation. 
The fraction of z $>$ 1 objects is
found to be $<$ 32\%, $<$ 43\% and $<$ 40\% of the extragalactic counts
 at 6.75$\mu$m, 15$\mu$m and 5 GHz, respectively.\\

Using deep photometry in rest-frame UV, visible, near-IR, mid-IR and radio,
we have derived spectral energy distributions (SEDs) which {\it samples
the wavelength range in which most of the energy of field galaxies is
emitted} (Figure 3). These have been compared to well-known local galaxy
 templates (from Schmitt et al, 1998). All the CFRS 1415+52 have been 
spectrally classified using this technics, and their FIR luminosities
 derived by interpolation. Our classification
allow us to estimate the AGN contribution, and then to remove it
when deriving the SFR density.\\

\small
\begin{figure}
  \begin{center}
    \leavevmode
    \psfig{file=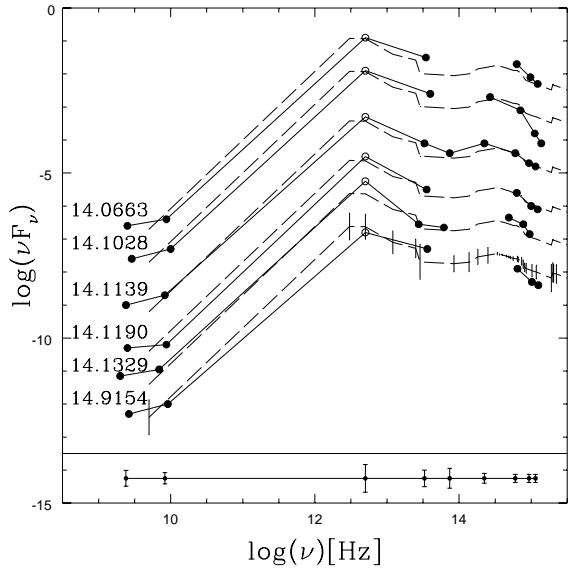,height=8cm,width=8cm} 
  \end{center}
  \caption{{\small Comparison of the spectral energy distribution 
  (SED) of 6 heavily reddened starbursts (filled circles, 
separated by arbitrary  vertical shifts) with a local 
  averaged starburst SED  (dashed line, from Schmitt  {\it et al.}, 1998). Fluxes 
  at 60$\mu$m (open circles) are derived from radio fluxes according
 to the radio-FIR 
  correlation ($S_{60\mu m}= 125 S_{5Ghz}$, Franceschini  {\it et al.}, 1994).
Superposed to the fit of 14.9154 are vertical bars which display at
each wavelength the standard deviation of the local template. The bottom panel
presents the standard deviation of the 6 starbursts.}}
 \label{figure3}
\end{figure}
\normalsize

We find that the SFR density derived by FIR fluxes is $\sim$ 2.3 higher
 than that
previously estimated from UV fluxes (Figure 4). No apparent changes 
with the redshift have been found within the range 0$\le$z$\le$1. The
 corresponding global
extinction is $A_{V}=0.55\pm0.12$, very similar to the Gallagher et al (1989)
average value for local irregulars. A subsample of sixteen 15$\mu$m galaxies observed by the HST indicates 
that about a third of the star formation hidden by dust is associated with 
interacting galaxies (Figure 5). One percent of
the CFRS galaxies are strong and heavily reddened starbursts (Figure 3) with
their SFR ranging from 120 to 330 $M_{\odot}yr^{-1}$ and they contribute
to 25\% of the SFR density up to z=1.\\

\small
\begin{figure}
  \begin{center}
    \leavevmode
    \psfig{file=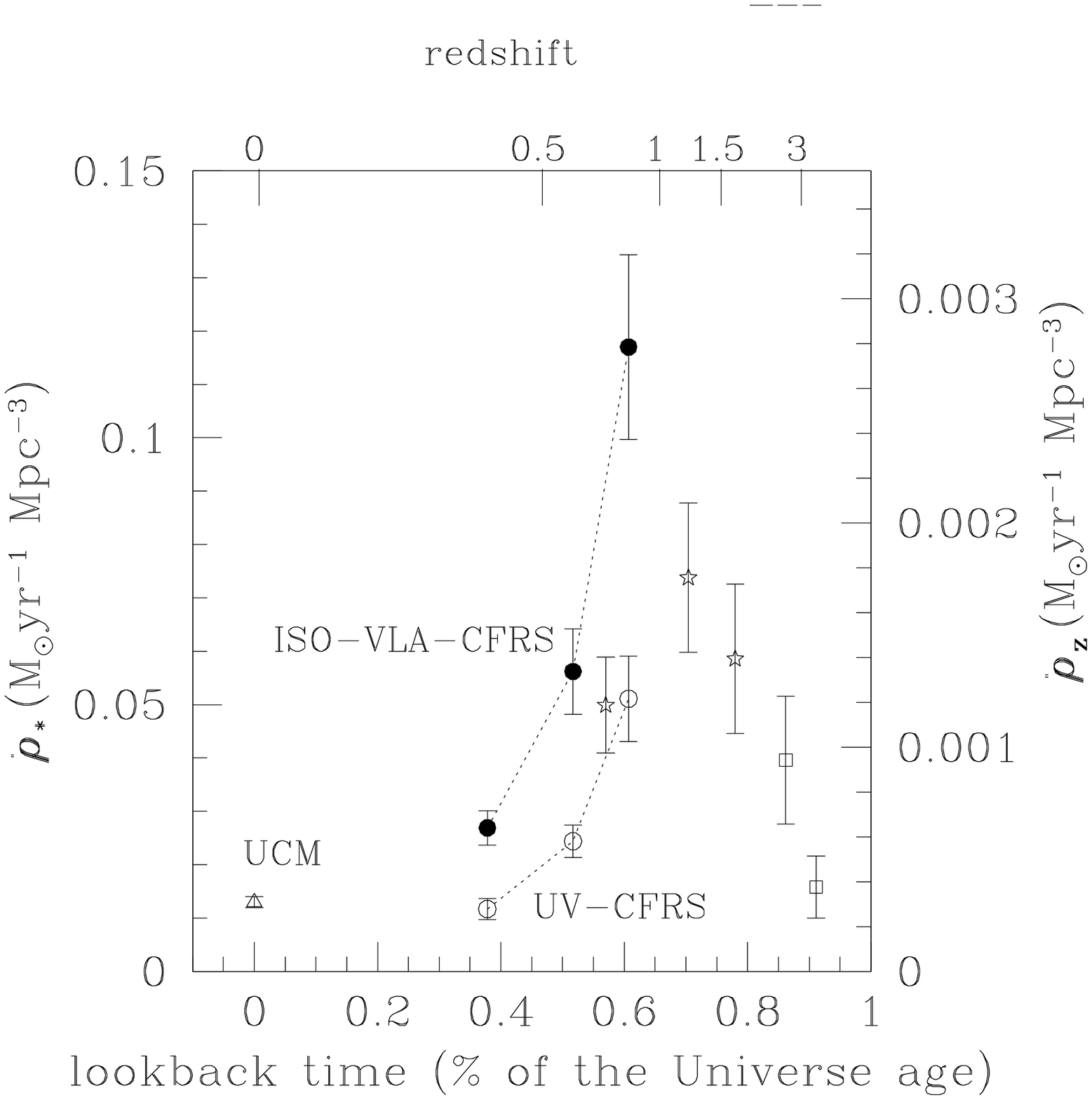,height=9.5cm,width=10cm} 
  \end{center}
  \caption{{\small Metal production and star formation history for 
$z\leq 1$ galaxies (see text). SFR estimates are assuming a Salpeter IMF from
0.1 $M_{\odot}$ to 100 $M_{\odot}$. Our
  points (filled circles, labeled ISO-VLA-CFRS) are 2.3 times higher 
in SFR density or in metal production than those 
(open circles) previously derived from 
the UV flux density at 2800\AA. Other points are 
  from Gallego {\it et al.} (1995, open triangle), Connolly{\it et al}
(1997) (open stars), Madau {\it et al.} 
(1998, HDF) (open squares), and have not been corrected for extinction.}}
 \label{figure4}
\end{figure}
\normalsize

\small
\begin{figure}
  \begin{center}
    \leavevmode
    \psfig{file=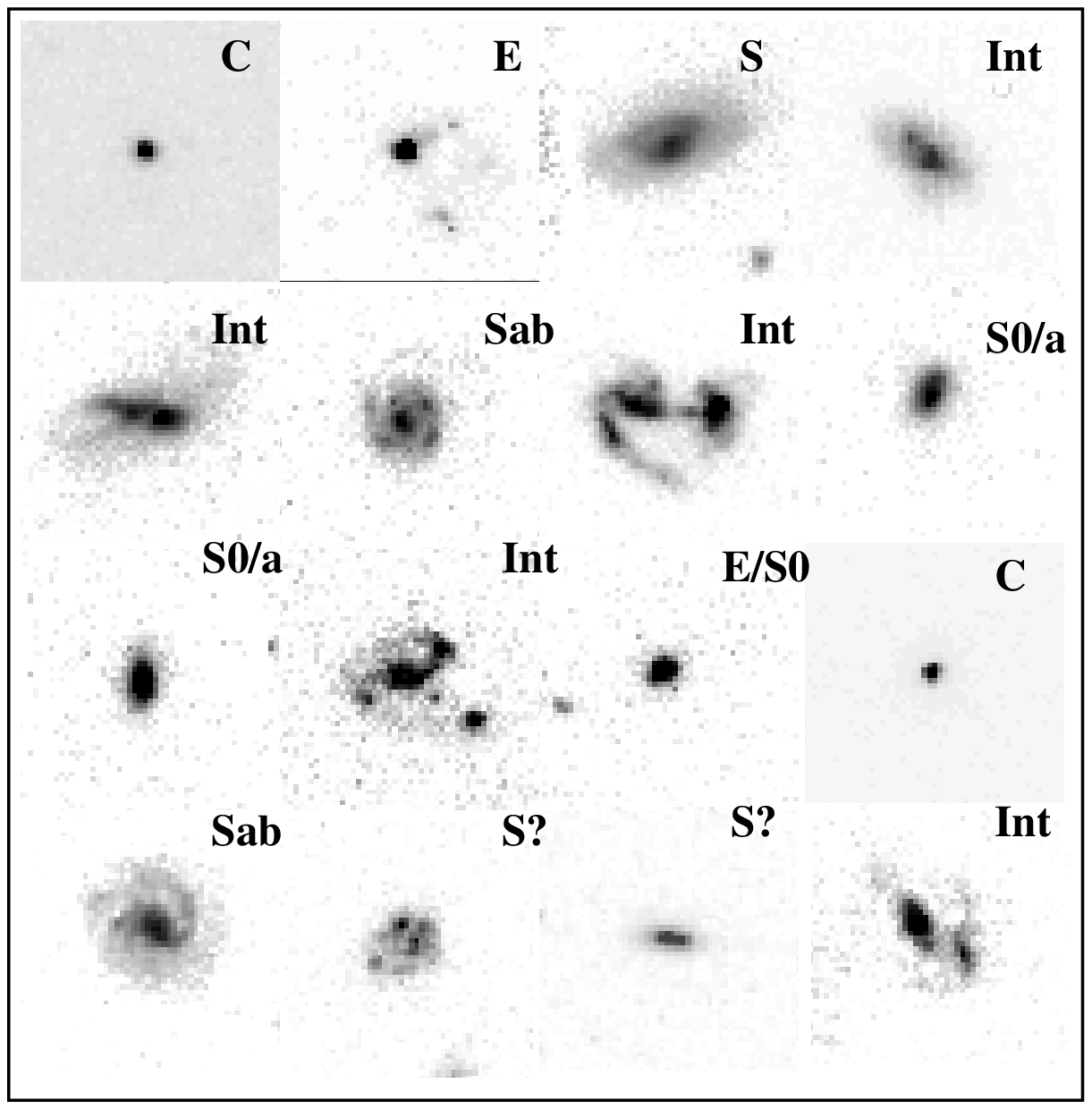,height=8cm,width=8cm} 
  \end{center}
  \caption{{\small HST 5''$\times$5'' images obtained with the F814W filter
  of 16 sources detected at 15$\mu$m with I$_{AB} < 22.5$. For each
  galaxy, the morphological classification is given in the upper right 
  corner.}}
 \label{figure5}
\end{figure}
\normalsize

These results provide SFR density values by $\sim$ 3.5 lower than those of
Rowan Robinson (1997) from the small HDF area. On the other hand they are
consistent with the Richards et al (1998) analysis who found only one strong starburst in the HDF. Nevertheless, our derived SFR density seems
too high, because the derived stellar mass formed since z=1 until
now is comparable or higher than the present day stellar mass. An
alternative would be that the most powerful starburst were preferentially
forming high mass stars, i.e. their IMFs are truncated at the low mass end.
 Another
possibility would be an underestimate of the local stellar mass density.\\
\\
Further results for an other CFRS field and spectroscopic follow 
up would provide a larger sample in order to test if extinction
properties show redshift evolution as well as to extend the predictions
 until z=1.5 near the possible peak of the SFR density. Deep SCUBA
 observations of some portions of these fields will be soonly available
(Eales et al and Lilly et al, 1998, in preparation).
   
\section{Morphologies from the HST}
\subsection{Observations and classifications}
During the last three years the CFRS team has merged their efforts with those
of the LDSS team, in order to reach significant amounts of observing time
 at the HST. About 250 of the 600 CFRS galaxies have been observed with
the WFPC2/F814W filter with integration times ranging from 4400 to 7400s.\\

Each galaxy has been analysed through two independent technics, including
systematic fits of the luminosity profile by combination of bulge 
($r^{1/4}$ law) and disk (exponential law) models, and a visual inspection
of images by three independent team members (see Schade et al, 1995 and Brinchman et al, 1998).  A morphological type has been given to each galaxy  following a scheme similar to the Hubble classification (see Figure 6).   
Morphological studies should account for the
shift in the rest-frame wavelength of the observation (for example the 
F814W filter samples the B band at z=1). This redshift dependent bias can
be very important since galaxies appear much less structured in the UV
than at redder wavelengths, and this can mimic an apparent evolution. By
 redshifting the Frei et al (1996) local galaxies it has been estimated
 that $\sim$ 24\%$\pm$11\% of the true spirals 
would have been classified as peculiars at z=0.9 (Brinchman et al, 1998).
All the numbers derived in the following have been computed after
 allowing for these biases.\\

Based on our classification, several subsamples of {\it morphologically}
selected objects can be drawn. These include disk galaxies (those with
bulge/total energy ratio smaller than B/T= 0.5), elliptical galaxies (B/T=1) and
compact galaxies. Determination of the scale length (for example the
exponential scale length of disks, $\alpha^{-1}$) is limited by the
pixel size of the HST, and is secure only for disks with
$\alpha^{-1}>3h_{50}^{-1}$kpc. The CFRS sample is well adapted for 
studying large systems (i.e. large disks 
and bulges) because it includes all galaxies from z=0.1 to z=1 with
$M_{B}(AB)\le$ -20.5.

\begin{figure}
  \begin{center}
    \leavevmode
    \psfig{file=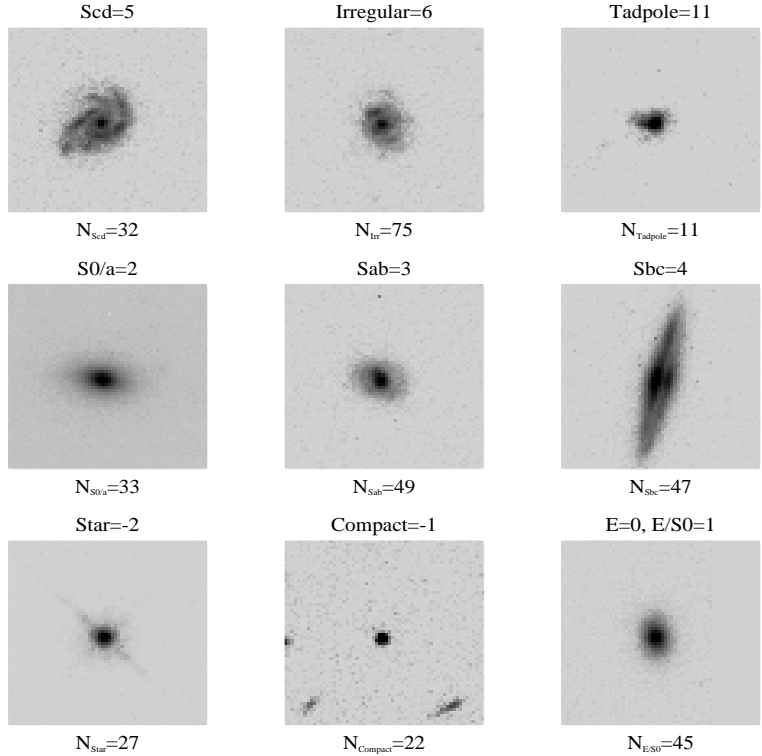,height=10cm,width=10cm} 
  \end{center}
  \caption{{\small Examples of the morphological types given to CFRS/LDSS
   galaxies,
    with the total number of objects in each class in the survey (from
    Brinchman et al, 1998).}}
  \label{figure6}
\end{figure}
\normalsize

\subsection{Large systems}
\subsubsection*{Large disks: $\alpha^{-1}\ge 4h_{50}^{-1}$kpc and B/T$\le$0.5}

Lilly et al (1998) gathered a subsample of 42 such galaxies from z=0.1 to z=1. 
Their number densities show no apparent change from z=0 to z=0.75, and are
consistent with a small decrease by $\le$30\% beyond z=0.75. Based
on UV and [OII]3727 fluxes, their star formation rates
at z$\sim$0.85 were 2.5-3 times higher on average than that at z=0. So the
large reported evolution is not mainly associated with large disks. An
interesting result shown by Lilly et al (1998) is that disk selected
galaxies had later type at higher redshift (median type is
Sab at z=0.375 compared to Scd at z=0.85).

\subsubsection*{Large ellipticals: $r^{1/4}$ luminosity profile and $r_{e}\ge 3h_{50}^{-1}$kpc}

30 such galaxies have been identified in the sample by Schade et al (1998, in
preparation) from z=0.2 to z=1. Despite of the small number, there 
are no obvious changes
in their number density from the low to the high redshift bins. This result
apparently contradicts the Kauffman et al (1996) analysis of the 
CFRS sample, who found a decrease in the
space density of galaxies redder than a rest-frame elliptical template.
This contradiction is simply due to the different definitions of ellipticals
by the two groups: a significant fraction of the
 spectrophotometric ellipticals of Kauffman et al (1996) are indeed disk
dominated galaxies (50\% of them have B/T$\le$0.5). 

\subsection{Irregular galaxies}
After allowing for redshift biases, Brinchman et al (1998) found a large
redshift increase of the fraction of irregular galaxies (from 9\% at z=0.375 to
32\% at z=0.85). This results is consistent with the migration of disk
galaxies towards late type at higher redshift. At z$ \sim $1, galaxies were
less regular than at present day.

\subsection{Compact galaxies}
It has been claimed that a considerable fraction of the star formation activity
seen at high redshift occurs in compact galaxies (Guzman et al, 1997). 
Compact galaxies in the HDF have been selected as having small half light radius
($r_{1/2}\le$ 0.5 arcsec). Half light radii have been also computed for the
CFRS galaxies, and Lilly et al (1998) found that large changes in
the galaxies population are due to systems with $r_{1/2}\le 5h_{50}^{-1}$kpc
(this corresponds to a disk scale $\alpha^{-1}\le 3h_{50}^{-1}$kpc).
We have also computed a compactness parameter which is based on the 
ratio of luminosities calculated within two different apertures 
(5 and 15$h_{50}^{-1}$kpc, respectively),
 and has been corrected for possible disk inclination (axis ratio b/a):   

\begin{equation}
compactness= I_{F814W}(5kpc)-I_{F814W}(15kpc)-2.5 log(b/a).
\label{eq:curv}
\end{equation}

This parameter correlates very well with the half radius and Figure 7 shows
that beyond z=0.5 (and especially beyond z=0.75), a significant number of
blue compact galaxies have very large UV luminosities.

\small
\begin{figure}
  \begin{center}
    \leavevmode
    \psfig{file=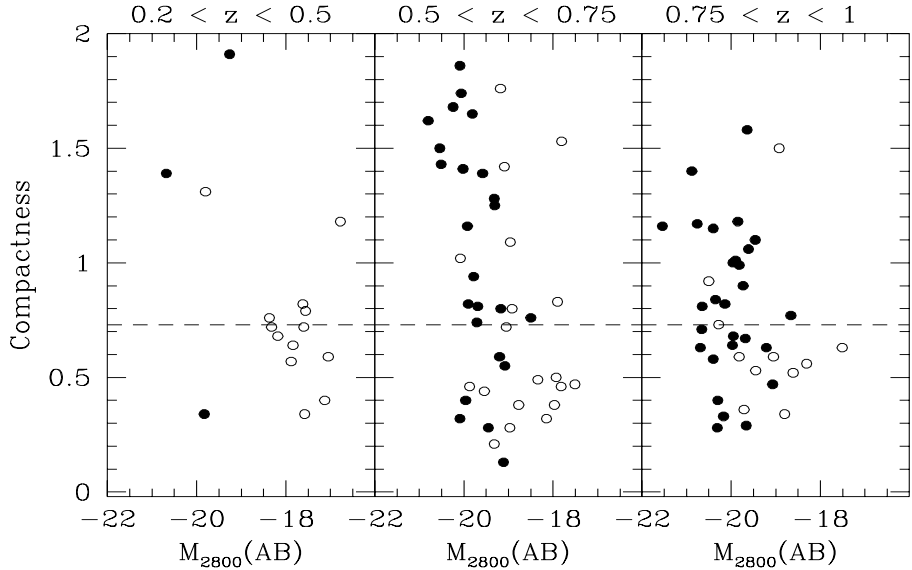,height=7.5cm,width=12cm} 
  \end{center}
  \caption{{\small Compactness parameter against $M_{2800}$ for the whole
    CFRS-HST sample (125 galaxies with $M_{B}(AB)\le$-20.5). Solid circles 
    distinguish galaxies bluer
    than a rest-frame Sbc. Three redshift bins are displayed and the
    two highest redshift bins represent almost an equipartition of the 
    comoving volume from z=0.5 to z=1. Compact
    galaxies (with $r_{1/2}\le 5h_{50}^{-1}$kpc) are those with 
    compactness$\le$0.73 (below the dashed line). Their contribution
 to the UV luminosity
    density increases very strongly with the redshift.}}
 \label{figure7}
\end{figure}
\normalsize
   
Figure 8 presents the HST/WFPC2/F814W images of the 29 blue compact galaxies. 
These galaxies are the most rapidly evolving population in the CFRS sample: 
 at z=0.85, their UV luminosity density was $\sim$ 15 times larger 
than at z=0, and they contribute to as high as 30\% of the global UV 
luminosity density. Of course these results could be modified if extinction
is accounted for. This is however a good confirmation of the Guzman et al
 (1995) result. On the other hand, the CFRS is sampling much brighter galaxies 
($M_{B}(AB)\le$-20.5) than those selected by Guzman et al in the HDF. 
This implies that the CFRS compact
galaxies at high redshift have sizes comparable to present-day dwarves (i.e.
disk scale smaller than 3$h_{50}^{-1}$kpc) while their blue luminosities are
10 to 100 times larger. Converted into star formation rates, these galaxies 
formed from 2 to 10 $M_{\odot}yr^{-1}$, a much higher rate than in present day 
dwarves. In the highest redshift bin (z$\sim$0.85), they span the whole 
range of morphological types. 

\begin{figure}
  \begin{center}
    \leavevmode
    \psfig{file=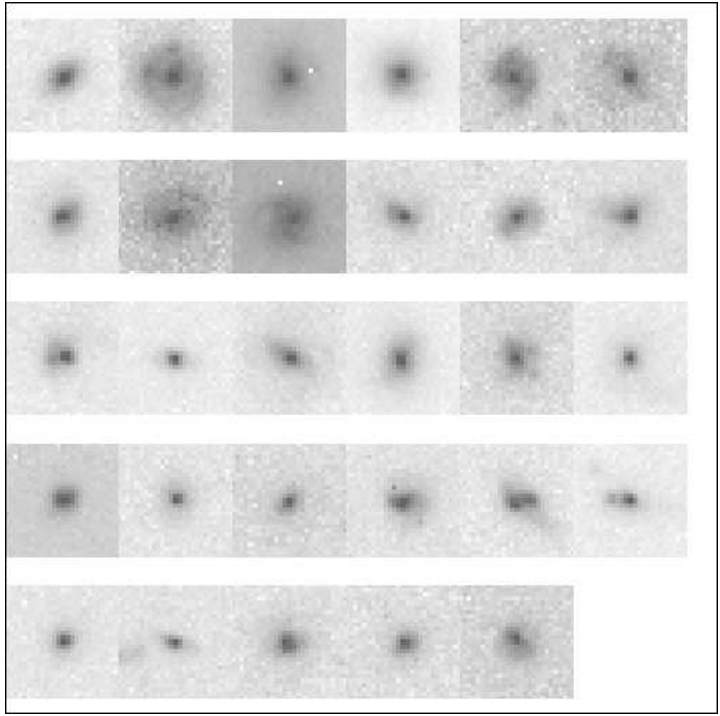,height=9.5cm,width=9.5cm} 
  \end{center}
  \caption{{\small HST images of the 29 blue compact galaxies with 
    $M_{B}(AB)\le$-20.5. Galaxies are ordered by increasing 
    redshift, from z=0.1 (top-left) to z=1.1 (bottom-down). Size of
    each stamp is 2.5". Some are interacting systems, some others are embedded 
in large and low surface brightness envelops.}}
 \label{figure8}
\end{figure}
\normalsize
   
\section{Conclusion}

The CFRS probes $\sim$ 10 Gyr lookback times, and the corresponding fields are
the best suited for follow up at other wavelengths. The
calibration of the SFR from UV luminosities is truly uncertain for 
individual objects, because of uncontrolled extinction effects. The calculation
of the SFR density is also affected by the AGN contribution to the UV
luminosity density ($\sim$ 30\%). A multiwavelength analysis of
field galaxies from UV to mid IR and hence to radio wavelength allow us to
derive FIR luminosities by interpolation. It provides an estimate of the SFR
which does not sensitively depend on the extinction. The global 
correction related to extinction is not as strong as previously reported
 by Rowan-Robinson et al (1997), 
and reaches factor of 2.3 until z=1. One percent of the CFRS galaxies are
strong and heavily reddened starbursts with SFRs from 120 to 330 
$M_{\odot}yr^{-1}$ and they contribute to 25\% of the SFR density. A third of the star formation detected at FIR is related to interacting systems. In
most of the galaxies spanning the whole range of SFRs there are evidences for 
an important A star population and hence for star formation occuring during
long periods of time ($\sim$ 1 Gyr).\\

The rapidly evolving population of compact/Irr galaxies contributed to
30\% of the SFR density at z$\sim$ 0.85. An important
problem is the "devenir" of these systems which
 produce substantial amount of stars in small volumes (SFR of
2-10 $M_{\odot}yr^{-1}$ within a radius $\le$
3$h_{50}^{-1}$kpc). One would like to know if large disks, still observed
 at z$\sim$ 1, are stable or formed and then destroyed,
 knowing that the interaction rate is rapidly increasing with the redshift.
 Studies of the dynamical formation
of galaxies require at an 8m telescope a 2D-spectrograph with
 high spectral resolutions, i.e. R$\ge$ 10000 to resolve 
dynamical elements separated by $\sim$ 10$kms^{-1}$.

\vfill
\end{document}